\documentclass[prl,twocolumn,tightenlines,floatfix]{revtex4}
\usepackage{graphicx,amsfonts,amssymb,amsmath}

\newcommand{\nc}{\newcommand}
\nc{\al}{\alpha}
\nc{\ga}{\gamma}
\nc{\de}{\delta}
\nc{\ep}{\epsilon}
\nc{\ze}{\zeta}
\nc{\et}{\eta}
\renewcommand{\th}{\theta}
\nc{\Th}{\Theta}
\nc{\ka}{\kappa}
\nc{\la}{\lambda}
\nc{\rh}{\rho}
\nc{\si}{\sigma}
\nc{\ta}{\tau}
\nc{\up}{\upsilon}
\nc{\ph}{\phi}
\nc{\ch}{\chi}
\nc{\ps}{\psi}
\nc{\om}{\omega}
\nc{\Ga}{\Gamma}
\nc{\De}{\Delta}
\nc{\La}{\Lambda}
\nc{\Si}{\Sigma}
\nc{\Up}{\Upsilon}
\nc{\Ph}{\Phi}
\nc{\Ps}{\Psi}
\nc{\Om}{\Omega}
\nc{\ptl}{\partial}
\nc{\del}{\nabla}
\nc{\be}{\begin{eqnarray}}
\nc{\ee}{\end{eqnarray}}
\nc{\ov}{\overline}
\nc{\gsl}{\not\!}

\newcommand{\bi}[1]{\bibitem{#1}}
\newcommand{\fr}[2]{\frac{#1}{#2}}

\begin{document}

\setcounter{page}{1}

\title{Theta-Induced Electric Dipole Moment of the Neutron via QCD Sum Rules\\ $\;\;$ \\}

\author{Maxim Pospelov and Adam Ritz}
\affiliation{Theoretical Physics Institute, School of Physics and Astronomy \\
         University of Minnesota, 116 Church St., Minneapolis, MN 55455, USA}

\begin{abstract}
Using the QCD sum rule approach, we calculate the electric dipole
moment of the neutron induced by a vacuum $\theta$--angle to
approximately 40-50\% precision, 
$d_n = 2.4\times 10^{-16}\bar \theta e\cdot cm$.
Combined with the new experimental bound, 
this translates into the limit $|\bar \theta|<3\times 10^{-10}$.
\end{abstract}

\maketitle

\newpage

The impressive experimental limits on the electric dipole moments (EDMs) of 
neutrons and heavy atoms in general put a very strong constraint on 
possible flavor-conserving CP-violation around the electroweak 
scale \cite{KL}. This precision means that EDMs can in principle
probe a high energy scale by limiting the coefficients of
operators with dimension$\geq 4$, such as $G\tilde{G}$,
$GG\tilde{G}$, and $\ov{q}G\si\ga_5 q$ etc.. However, in practice, while
these operators can be perturbatively evolved down to a scale of order  
$1$ GeV, the ultimate connection 
between high energy parameters and low energy EDM observables 
necessarily involves non-perturbative physics.

The $\th$--term, $\th G\tilde{G}$, which has dimension=4 may be interpreted
as the lowest dimension CP-violating operator,
un-suppressed by the high energy scale. 
Experimental tests of CP symmetry suggest
that $\th$ is small and, among different CP-violating
observables, the EDM of the neutron ($d_n(\th)$) is the most sensitive to its 
value \cite{nEDM,nEDM1}. However, the calculation of 
$d_n(\th)$ is a long standing 
problem \cite{baluni,CDVW,AH}. According to Ref.~\cite{CDVW}, 
an estimate can be obtained within chiral perturbation 
theory, relying on the numerical dominance of contributions
proportional to $\ln{m_\pi}$ near the chiral limit. 
However, there are also non-logaritmic contributions, incalculable 
within this formalism, which in principle can be numerically 
more important than the logarithmic piece 
away from the chiral limit. As a consequence one is unable to estimate 
the uncertainty of the prediction \cite{CDVW}.

There are a number of important incentives for refining the calculation of 
$d_n(\theta)$. While $\th$ primarily arises as a fundamental vacuum
parameter, one may also induce calculable corrections via the
integration over heavy fields, and such corrections are significant
in theories where the fundamental parameter is absent. One scenario of this
type arises as a result of exact P or CP symmetries \cite{P}, whose subsequent
spontaneous breakdown allows radiative corrections to induce an
effective $\th$--parameter, which is then the dominant source for $d_n$.
Alternatively, the fundamental $\th$ parameter may be removed 
via the axion mechanism, but then a small but finite effective $\th$--angle 
will again survive, induced by the linear shift of the axionic potential 
due to higher-dimensional CP-odd operators
\cite{BUP}. In this case, the calculation of $d_n(\theta)$ 
is part of a more generic calculation of $d_n$. 

In this letter, we apply the QCD sum rule method \cite{sr} to obtain
an estimate for $d_n(\th)$ beyond chiral perturbation theory.
Within currently available analytic techniques, QCD sum rules 
appears the most promising approach to this problem
as it has, in particular, been used 
successfully in the calculation of certain baryonic 
electromagnetic form factors \cite{is,BY}. 
Within the sum rule formalism, physical properties of the 
hadronic resonances are expressed via a combination of
perturbative and nonperturbative contributions, the latter parametrized in 
terms of vacuum quark and gluon condensates. We note that previously 
QCD sum rules were used to estimate the neutron EDM induced by a CP-odd color 
electric dipole moment of quarks \cite{KKY,kw} . Surprisingly, the results 
disagree with the naive estimates based on the chiral loop approach. 
The calculation of $d_n(\theta)$ using QCD sum rules will certainly 
help to resolve this controversy. 

The approach we shall use follows recent work \cite{PR} on
the $\th$-induced $\rho$--meson EDM in reducing the operator 
product expansion to a set of vacuum 
condensates taken in an electromagnetic and topologically nontrivial 
background. Expansion to first order in $\theta$ results in the appearance
of matrix elements which can be calculated via the use of current 
algebra \cite{C,svz}. In this approach the $\theta$--dependence 
arises with the correct quark mass dependence, and the relation 
to the U(1) problem becomes explicit as $d_n(\theta)$ vanishes when the 
mass of the U(1) ``Goldstone boson'' is set equal to the mass of pion. 

The starting point for the calculation is 
the correlator of currents $\et_n(x)$ with quantum
numbers of the neutron in a background with nonzero
$\th$ and an electromagnetic field $F_{\mu\nu}$,
\be
 \Pi(Q^2) & = & i\int d^4x e^{iq\cdot x}
    \langle 0|T\{\et_n(x)\ov{\et}_n(0)\}|0\rangle_{\th,F},
     \label{pi}
\ee
where $Q^2=-q^2$, with $q$ the current momentum.

Firstly, it is important to note that if 
CP-symmetry is broken by a generic quark-gluon CP-violating source
($\theta$--term in our case), 
the coupling between the physical state (neutron) described by a spinor $v$ 
and the current $\eta_n$ acquires an additional phase factor
\be
\langle 0|\et_n|N\rangle=\lambda U_{\al}v, \;\;\;\;\;\;\;\;\;\;
     U_{\al}=e^{i\al\ga_5/2}. \label{coupling}
\ee
The existence of this unphysical phase $\al$ is apparent already when
one considers the sum rule for the neutron 
 mass, which in the 
absence of CP-invariance can have an additional Dirac structure
proportional to $i\gamma_5$. When we turn to electromagnetic form factors, this
angle can mix electric ($d$) and magnetic ($\mu$) dipole moment structures 

\begin{figure}
\includegraphics[width=8.5cm]{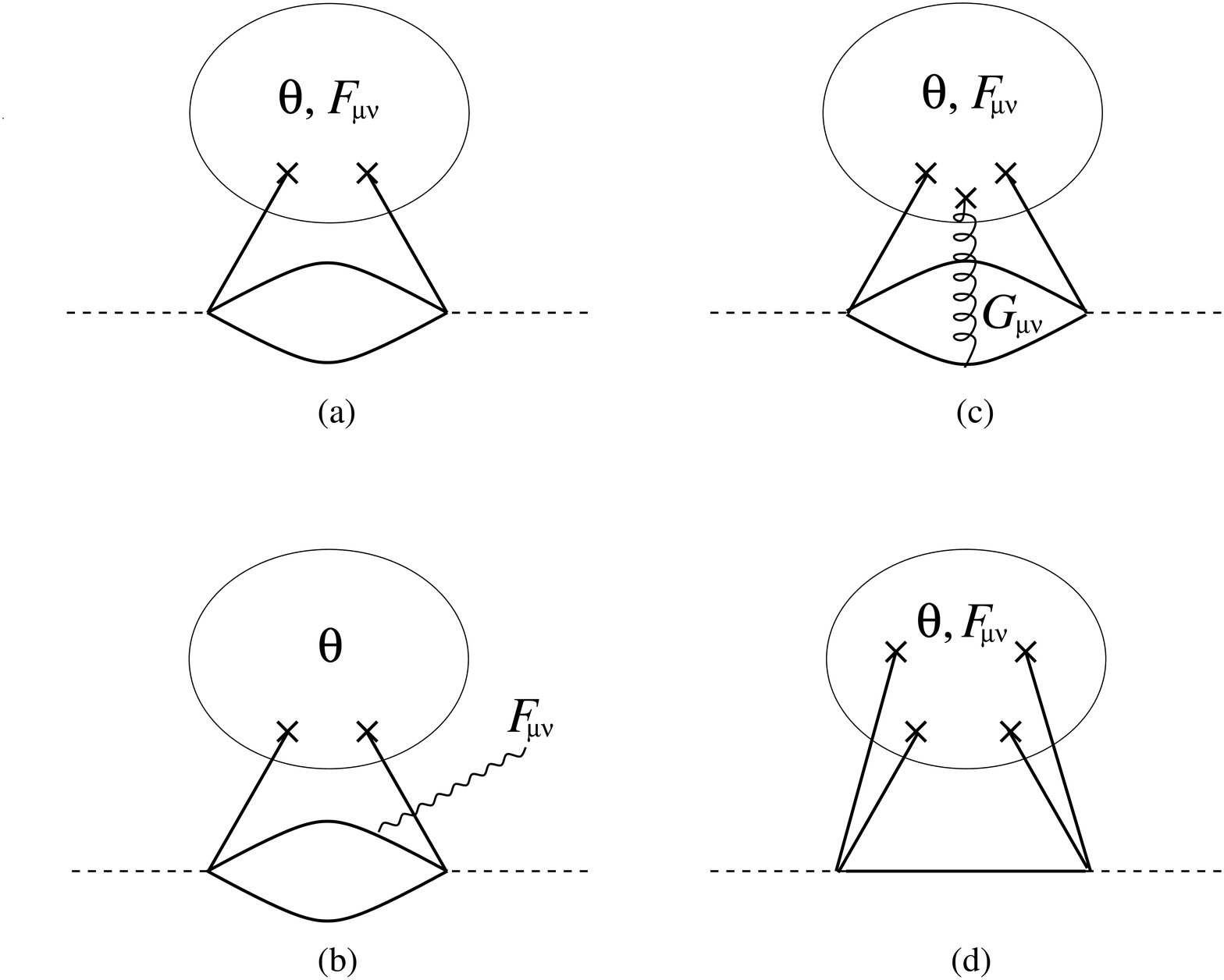}
 \caption{\footnotesize Various contributions to the CP-odd structure
  $\{\tilde{F}\si,\gsl{q}\}$. (a) is the leading order contribution
while (b), (c) and (d) contribute at subleading order.}
\end{figure}

\noindent
and
complicate the extraction of $d$ from the sum rule.
Consequently, in general the neutron double pole contribution on the 
phenomenological side of the sum rule will be proportional to the 
following expression
\be
 U_{\al}(\gsl{q}+m_n)(\mu F\si-d\tilde F\si)(\gsl{q}+m_n)U_{\al} &&
    \nonumber \\
 && \!\!\!\!\!\!\!\!\!\!\!\!\!\!\!\!\!\!\!\!\!\!\!\!\!\!\!\!\!\!\!\!\!\!\!\!
   \!\!\!\!\!\!\!\!\!\!\!\!\!\!\!\!\!\!\!\!\!\!\!\!\!\!\!\!\!\!\!\!\!\!\!\!
     \!\!\!\!\!\!\!\!
    = m_n\{\mu F\si-d\tilde F\si,\gsl{q}\}  \nonumber\\
 && \!\!\!\!\!\!\!\!\!\!\!\!\!\!\!\!\!\!\!\!\!\!\!\!\!\!\!\!\!\!\!\!\!\!\!\!
    \!\!\!\!\!\!\!\!\!\!\!\!\!\!\!\!\!\!\!\!\!\!\!\!\!\!\!\!\!\!\!\!\!\!\!\!
    \!\!\!\!\!\!\!\!\!\!\!\!\!\!\!\!\!\!\!\!\!\!\!\!\!\!\!\!\!\!\!\!\!\!\!\!   
    + \sum(\mu\pm \al d){\cal O}(F) + \sum(d\pm \al\mu){\cal O}(\tilde{F}),
\ee  
in which ${\cal O}(F)$ and ${\cal O}(\tilde{F})$ are operators depending
on the electromagnetic field strength $F$ and its dual $\tilde{F}$
which contain an even number of $\ga$--matrices. We see that only 
Lorentz structures with an odd number of 
$\gamma$-matrices are independent of $\al$. In calculating $d_n$, it is then
clear that we should study the operator $\{\tilde{F}\si,\gsl\!q\}$, 
as this is the unique choice with a coefficient independent of $\al$.

The interpolating current $\et_n$ is conveniently parametrised in the form,
\be
 \et_n & = & j_1 +\beta j_2,
\ee
where the two contributions are given by
\be
 j_1 & = & 2\ep_{abc}(d_a^TC\ga_5u_b)d_c \\
 j_2 & = & 2\ep_{abc}(d_a^TCu_b)\ga_5d_c.
\ee
$j_2$ vanishes in the nonrelativistic limit, and lattice simulations
have shown that $j_1$ indeed provides the dominant projection onto
the neutron (see e.g. \cite{leinweber}). 
From (\ref{coupling}), we may define 
the coupling to the neutron state in the form 
$\langle 0|\et_n|N\rangle=(\la_1+\beta\la_2)v$.

Within the sum rules formalism, one has the imperative
of suppressing the contribution of excited states and higher dimensional
operators in the operator product expansion (OPE), and thus 
its convenient to choose $\beta$ to this end. 
Ioffe has shown 
that $\beta=-1$ is an apparently optimal choice for the mass sum rule.
However, it is clear that this optimization may differ for different
physical observables. We shall therefore keep $\beta$ arbitrary, and 
optimize once we have knowledge of the structure of the sum rule.

We now proceed to study the OPE associated with (\ref{pi}). 
The relevant diagrams we need to consider are shown in Fig.~1 
((a), (b) and (c)). Diagrams of the form (d), although suffering
no loop factor suppression, are nonetheless suppressed due to 
combinatorial factors
and the small numerical size of $(\langle \ov{q}q\rangle)^2$.  

In parametrizing $\th$, we shall take a general initial condition
in which a chiral rotation has been used to
generate a $\ga_5$--mass, so that
\be
 {\cal L} & \sim & \cdots -\th_q m_* \sum_f\ov{q}_fi\ga_5 q_f
       +\th_G\frac{\al_s}{8\pi}G^a_{\mu\nu}\tilde{G}^a_{\mu\nu}+\cdots,\;\;\;\;
          \label{general}
\ee
in which we restrict to $q_f=u,d$, and so $m_*=m_um_d/(m_u+m_d)$. The
physical parameter is of course $\overline{\th}=\th_q+\th_G$, but we 
shall keep the general form (\ref{general}) and calculate the OPE 
as a function of {\em both} phases. The independence of the final answer
of $\th_q-\th_G$ will provide a nontrivial check on the consistency
of our approach. We shall find that this requires the consideration of 
mixing with additional currents, a point we shall come to shortly.
 
The vacuum structure is conveniently encoded in a generalised propagator
expanded in the background field and the associated condensates.
We work as usual with a constant background electromagnetic
field, so that $A_{\mu}(x)=-\fr{1}{2}F_{\mu\nu}(0)x^{\nu}$, and
use a fixed point gauge for the gluon potential,
$A_{\mu}^at^a(x)=-\fr{1}{2}G_{\mu\nu}^a(0)t^ax^{\nu}$.
The electromagnetic field
dependence is determined in terms of the magnetic susceptibilities $\ch$,
$\ka$ and $\xi$, defined as \cite{is}:
\be
 \langle 0| \ov{q}\si_{\mu\nu}q|0\rangle_F & = & \ch_q F_{\mu\nu}
           \langle 0| \ov{q}q|0\rangle \nonumber\\
  g\langle 0| \ov{q}(G_{\la\si}^at^a)q|0\rangle_F 
   & = & \ka_q F_{\mu\nu}\langle 0| \ov{q}q|0\rangle \nonumber\\
 2g\langle 0| \ov{q}\ga_5(\tilde{G}_{\la\si}^at^a)q|0\rangle_F 
   & = & i\xi_q F_{\mu\nu}\langle 0| \ov{q}q|0\rangle \nonumber,
\ee
while the $\th$--dependence is either explicit in the case of $\th_q$,
or extracted via use of the anomalous
Ward identity (see e.g. \cite{svz}) in the case of $\th_G$. 
The $\theta$--dependence of the OPE then follows as a consequence of
$m_{\et}\gg m_{\pi}$, and disappears when $U(1)$-symmetry 
is restored ($m_\eta\rightarrow m_\pi$) \cite{PR}.  

Defining $iS(q)\equiv \langle 0|q_a(x) \ov{q}_b(0)|0\rangle_{F,\th}$,
and ignoring a trivial $\de$--function over colour indices, the 
leading order propagator adapted to the CP-odd sector
and appropriate for Fig.~1(a), takes the form,
\be
 S_{LO}(x) & = & \frac{\gsl x_{ab}}{2\pi^2x^4}
       -\frac{m_*}{4\pi^2x^2}(1-i\th_q\ga_5)_{ab} \nonumber \\
    && \!\!\!\!\!\!\!\!\!\!\!\!\!\!\!\!\!\!\!\! 
   -\frac{\tilde{\ch}_qm_*\ov{\th}}{24}F_{\al\beta}x_{\al}
   (\ga_{\beta}\ga_5)_{ba}
       -\frac{\tilde{\ch}_q}{24}(F\si(1+i\th_G\ga_5))_{ba}, \;\;\;\;\;
      \label{prop}
\ee
where we have denoted $\tilde{\ch}_q=\ch_q\langle \ov{q}q\rangle$, and we
shall henceforth follow \cite{is} and assume that $\ch_q=\ch e_q$ etc.,
with flavour independent parameters $\ch,\ka,\xi$. 

Diagrams (b) and (c) in Fig.~1
require, in addition, the leading order expansion in the background gluon and
electromagnetic fields, and $S_{NLO}$ is consequently more involved.
However, since the approach is analogous to that for the leading order terms,
we shall defer full details \cite{pr2}, and simply
present the final result for the OPE structure arising from
these diagrams. In momentum space we find
\be
 \Pi(Q^2) \!&=&\! -\frac{\ov{\th}m_*}{64\pi^2}\langle \ov{q}q\rangle
      \{\tilde{F}\si,\gsl q\}
   \left[\ch(\beta+1)^2(4e_d-e_u)\ln \frac{\La^2}{Q^2}\right.\nonumber\\
  && \!\!\!\!\!\!\!\!\!\!\!\!\!\!\!\!\!\!\!\!\!\!
    -4(\beta-1)^2e_d\left(1+\frac{1}{4}(2\ka+\xi)\right)
     \left(\ln\frac{Q^2}{\mu_{IR}^2}
       -1\right)\frac{1}{Q^2} \nonumber \\
  &&\!\!\!\!\!\!\!\!\!\!\!\!\!\!\!\!\!\!\!\!\!\!\!\!\!\!\!
 \left. - \frac{\xi}{2}\left((4\beta^2-4\beta+2)e_d+(3\beta^2+2\beta+1)e_u
     \right)\frac{1}{Q^2}\cdots\right],\;\;\;\;\;\; \label{ope}
\ee
where the first, second, and third lines correspond to contributions
from diagrams (a), (b)+(c), and (c), in Fig.~1, and $\mu_{IR}$ is an infrared
cutoff. The diagrams (d) which contribute at subleading order in 
$q^2$ are more problematic because they involve 
correlators not calculable in the chiral approach, e.g.
\be
\Pi_{(d)}\sim  
\int d^4x e^{iq\cdot x}
    \langle 0|T\{(\bar qq)^2(x),m_*\sum_f\ov{q}_fi\ga_5 q_f
 \}|0\rangle. \;\;\;
\ee
We estimate such contributions via saturation with the physical 
$\eta$ meson. The result turns out to be parametrically 
smaller than any term listed in
Eq. (\ref{ope}). 

At different stages of the calculation we observe 
the appearance of the unphysical phase $\th_G-\th_q$ which is closely related 
with the non-invariance of the currents $j_1$ and $j_2$ under a
chiral transformation. This invariance is, in fact, restored when we consider
the mixing of $j_1$ and $j_2$ with another set of currents, 
$i_1=2\ep_{abc}(d_a^TCu_b)d_c$ and $i_2=2\ep_{abc}(d_a^T\ga_5 Cu_b)\ga_5d_c$.
The mixing between these two sets is explicitly proportional to 
$\th_G-\th_q$. 
When properly diagonalized on the sum rule for $\gsl{q}$, the 
linear combination of these two sets of currents provides additional 
contributions proportional to $\th_G-\th_q$, exactly cancelling the unphysical 
piece of $\Pi(Q^2)$. Consequently, it is equivalent to take 
$\th_q=\th_G$ as the most convenient
basis when working with $j_1$ and $j_2$ 
where the unwanted phase, and the mixing between 
$j_{1,2}$ and $i_{1,2}$, are simply absent. This situation resembles 
the problem in obtaining the ``correct'' quark mass behavior for 
the EDM of $\rho$, addressed in detail in \cite{PR}. 

Inspection of the expression (\ref{ope}) shows that the standard choice of
$\beta=-1$ appropriate for CP-even sum rules will not be useful here
as it removes the leading order contribution. 
In general there are two motivated techniques for fixing the
mixing parameter $\beta$: (1) at a local extremum; or (2) to minimize 
the effects of the continuum and higher dimensional operators.
We find in this case that extremizing in
the parameter $\beta$ also leads to the unappealing cancelation
of the leading order contribution. Thus the 
most natural procedure appears to be to choose $\beta$ in order to cancel
the subleading infrared logarithm which is ambiguous as a result of 
the required infrared cutoff. This procedure actually 
mimics the original motivation for
$\beta=-1$ in the CP-even case. We then take $\beta=1$, and it is
this choice that we shall now contrast with the phenomenological side 
of the sum rule. It is worth to remarking, however, that
use of the ``lattice'' current with $\beta=0$ will also
produce a numerically similar result. 

On the phenomenological side of the sum rule we have
\be
 \Pi^{phen} &=&\frac{1}{2}\{\tilde{F}\si,\gsl q\}\left(
\fr{\lambda^2d_nm_n}{(q^2-m_n^2)^2} +
\fr{A}{q^2-m_n^2}\cdots\right). \;\;\;\;\;\;\;
          \label{phenfull}
\ee
We retain here double and single pole contributions,
the latter corresponding to transitions between the neutron and excited
states, but the exponentially suppressed continuum contribution will be 
ignored as it is not, in this case, sign definite.
Here we use $\la=\la_1+\beta\la_2$ and $A$ is an effective constant
parametrizing the single pole contributions.

After a Borel transform of (\ref{ope}) and (\ref{phenfull}) and
using $\beta=1$ as discussed earlier, 
we obtain the sum rule
\be
 \la^2 m_nd_n+AM^2 & = & -\frac{M^4}{32\pi^2}\ov{\th}m_*\langle \ov{q}q\rangle
   e^{m_n^2/M^2}\nonumber\\
   && \!\!\!\!\!\!\!\!\!\!\!\!\!\!\!\!\!\!\!\!\!\!\!\!\!\!\!\!\!
  \times\left[4\ch(4e_d-e_u)-\frac{1}{2M^2}\xi(4e_d+8e_u)\right]. \;\;\;\;
   \label{sumrule}
\ee  

The coupling $\la$ present in (\ref{sumrule}) may be obtained
from the well known sum rules for the tensor structures {\bf 1} and $\gsl{q}$ 
in the CP even sector (see e.g. \cite{leinweber} for a recent review).
We shall construct two sum rules in this way.
\begin{itemize}
\item {\bf (a)} Firstly, we extract a numerical value for $\la$ via a direct
analysis of the CP even sum rules. This analysis has been discussed before
and will not be reproduced here (see e.g. \cite{leinweber}). 
One uses $\beta=-1$, and obtains $(2\pi)^4\la\sim 1.05\pm 0.1$.
\item {\bf (b)} As an alternative,
we extract $\la$ explicitly as a function of $\beta$ 
from the CP-even sum rule for $\gsl{q}$, and substitute the result
into (\ref{sumrule}) choosing $\beta=1$. 
\end{itemize}

The conventional approach which we shall adopt here 
is to assume that $A$ is independent of $M$, and thus
the left hand side of (\ref{sumrule}) is linear in $M^2$ provided that
$\la$ is constant in the appropriate region for the Borel parameter. The
latter point in case (b) above may be checked explicitly. 

It is convenient to define an additional function $\nu(M^2)$,
\be 
 \nu(M^2) & \equiv & \frac{1}{\ov{2\th}m_*}
    \left(d_n+\frac{AM^2}{\la^2m_n}\right),
\ee
which is then determined by the right hand side of (\ref{sumrule}).

\begin{figure}
\includegraphics[width=8cm]{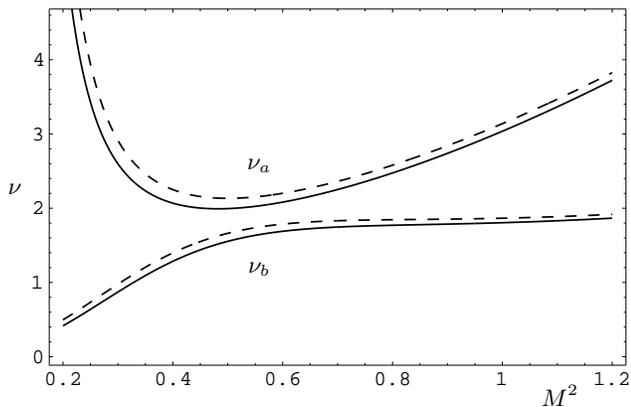}
 
\vspace{-3cm} \hspace{-8.5cm} $\nu$

\vspace{-0.7cm} \hspace{-2cm} $\nu_a$

\vspace{1cm} \hspace{-2cm} $\nu_b$

\vspace{1.4cm} \hspace{6cm} $M^2$

 \caption{\footnotesize The neutron EDM function $\nu(M^2($GeV$^2))$
is plotted according to the sum rules (a) and (b).
The dashed line shows the contribution from the leading order term
only.}
\end{figure}

The two sum rules described above for $\nu_a$ and
$\nu_b$ are plotted in Fig.~2, where the effect of the
higher dimensional terms in (\ref{sumrule}) proportional to
$\xi$ is also displayed. $\nu(M^2)$ is to be interpreted as
a tangent to the curves in Fig.~2.
For numerical calculation we make use of the following
parameter values: For the quark condensate, we take
$\langle 0|\ov{q}q|0\rangle = - (0.225\mbox{ GeV})^3$,
while for the condensate susceptibilities, we have the values 
$\ch = - 5.7 \pm 0.6 \mbox{ GeV}^{-2}$ \cite{chival}, and
$\xi = - 0.74 \pm 0.2$ \cite{kw}.
Note that $\ch$, which enters at $O(1/M^2)$, since it is dimensionful,
is numerically significantly larger than $\xi$. In extracting $\la$
in case (b) we also set a relatively large continuum threshold
$s_0=(2\mbox{GeV})^2$ for consistency with the CP-odd sum rule in which
this continuum is ignored.

One observes that both
sum rules have extrema consistent to $\sim 50\%$, suggesting that our
procedure for fixing the parameter $\beta$ is appropriate. Furthermore,
the differing behaviour away from the extrema implies that for consistency
we must assume $A$ to be small. One then finds 
$d_n\sim 2\bar\theta m_* \nu(M^2\sim 0.5$GeV$^2)$. It is also
interesting that the effective scale is around $M\sim 0.7$GeV which
is well below $m_n$, and should be cause for concern regarding the
convergence of the OPE. Nonetheless, one sees that the corrections
associated with the leading higher dimensional operators are still
quite small. This low scale is also the reason we have ignored
leading-log estimates for the anomalous dimensions, as their
status is unclear at this scale. A naive application leads to a small
correction that we shall subsume into our error estimate.

Extracting a numerical estimate for $d_n$ from Fig.~2, 
and determining an approximate error, we find the result\footnote{(04/2005) v3: This updated 
expression corrects an overall factor of two error present in previous versions.} 
\be
 d_{n} &=& \frac{(1.0 \pm 0.3)}{({\rm 500 MeV})^2}\,\ov{\th}\,e\,m_*,
\label{final}
\ee
for the neutron EDM, for which the dominant contribution naturally arises
from $\ch$.

Comparison with the result of ref. \cite{CDVW} indicates rather good
agreement in magnitude, due essentially to the low effective
mass scale $M\sim 700$MeV. The relation between the
calculation presented in this letter and the chiral logarithm 
estimate will be discussed in more detail
elsewhere \cite{pr2}, but we note in passing that the sum rule result
is parametrically enhanced ($O(1/N)$ vs. $O(1/N^2)$) at large $N_c$.

Combining our result with the recently improved experimental bound on $d_n$
\cite{nEDM1} we derive, allowing a generous 50\% uncertainty,
the limit on theta:
\be
|\bar \theta|< 3\times 10^{-10},
\ee
which is quite close to previous bounds.

In conclusion, we have presented a QCD sum rules calculation of 
the $\theta$-induced neutron EDM. This result is explicitly tied
to a set of vacuum correlators which are non-vanishing only in the absence
of a U(1) ``Goldstone boson''. The use of QCD sum rules in the chirally 
invariant channel allowed us to unambiguously extract $d_n(\th)$, and 
independence of the answer from any particular representation of the theta
term (\ref{general}) was checked explicitly.

{\bf Acknowledgments}
We would like to thank M. Shifman and A. Vainshtein for valuable discussions
and comments.

\bibliographystyle{prsty}

\end{document}